\documentclass[st,a4paper,twocolumn]{jpsj3}
\usepackage{txfonts}
\usepackage{bm}
\usepackage{graphicx}

\newcommand{\be}{\begin{eqnarray}}
\newcommand{\ee}{\end{eqnarray}}
\newcommand{\ba}{\begin{array}}
\newcommand{\ea}{\end{array}}
\newcommand{\no}{\nonumber}
\newcommand{\diff}{\mathrm d}
\newcommand{\e}{\mathrm e}

\title{Hamiltonian engineering for adiabatic quantum computation: \\
Lessons from shortcuts to adiabaticity}

\author{Kazutaka Takahashi}
\inst{Department of Physics, Tokyo Institute of Technology, 
Tokyo 152--8551, Japan}

\abst{
We discuss applications of shortcuts to adiabaticity (STA) 
to adiabatic quantum computation.
After reviewing the fundamental properties and the present status of STA
from the author's personal point of view, 
we apply the method to the adiabatic algorithm of the Grover's problem.
We discuss two possible implementations of STA for 
the adiabatic quantum computations: 
the method of quantum adiabatic brachistochrone and
the Lewis--Riesenfeld invariant-based inverse engineering.
}

\begin{document}
\maketitle

%%%%%%%%%%%%%%%%%%%%%%%%%%%%%%%%%%%%%%%%%%%%%%%%%%%%%%%%%%%%%%%%%%%%%%%%%%%%%%%%%
\section{Introduction}

Adiabatic quantum computation (AQC) is one of the computational methods 
to solve optimization problems using 
quantum effects~\cite{ACF, Somorjai, AHS, FGSSD, KN, BBRA, FGGS, FGGLLP}.
We set the solution of the optimization problem 
to the ground state of the problem Hamiltonian $\hat{H}_{\rm p}$ 
and consider the time evolution with the Hamiltonian 
\be
 \hat{H}(t)=A(t)\hat{H}_{\rm d}
 +B(t)\hat{H}_{\rm p}, \label{haqc}
\ee
where $\hat{H}_{\rm d}$ represents a driver Hamiltonian. 
The coefficients satisfy the initial condition $B(0)=0$ 
and the final condition $A(t_{\rm f})=0$ at the run time $t_{\rm f}$.
When we consider the time evolution starting from the trivial ground state 
determined by the driver Hamiltonian, 
the state reaches, roughly speaking, the nontrivial ground state 
of the problem Hamiltonian at $t=t_{\rm f}$.
By measuring the final state, we obtain the solution of the optimization problem.

Since AQC uses the Schr\"odinger equation,
we can exploit physical intuitions and various techniques 
developed for physical systems.
The property that the final state becomes the ground state of the problem 
Hamiltonian is guaranteed by the adiabatic theorem~\cite{BF, Kato}.
We require an infinitely-slow variation of the Hamiltonian to apply the theorem.
In actual situations, it is an approximation rather than the theorem.
The rigorous condition that the adiabatic approximation gives 
the correct result was studied in various works~\cite{JRS, LRH, AL18}, 
which show that establishing the rigorous condition is not so simple 
and is a cumbersome task.
In addition to that, the choices of the driver Hamiltonian and
the schedule $(A(t), B(t))$ are completely arbitrary, 
putting aside restrictions in experiments.
The only property we need is that $\hat{H}_{\rm p}$ and $\hat{H}_{\rm d}$ do not 
commute with each other.

Here, we use the word AQC, rather than quantum annealing~\cite{FGSSD, KN}.
This is because we want to study closed systems in the present paper.
In quantum annealing the system is generally coupled to the environment and 
we sometimes exploit dissipation and decoherence effects for a sampling.
Even though we restrict our analysis to the closed systems, 
we have not fully understood the mechanism of AQC. 
The theory is mainly based on the static picture and, as we mentioned above,
we have the arbitrariness of choosing the Hamiltonian.
There is no general guiding principle to improve the performance.

In this paper, we discuss shortcuts to adiabaticity (STA)~\cite{STA} 
as a tool to understand the fundamental properties of AQC, 
and to optimize the algorithm.
This method treats the time evolution of the quantum states with finite speed.
However, we show that the adiabatic picture is still useful to understand 
the nonadiabatic time evolutions. 
It is not a contradiction and 
we discuss that STA can be a key 
to understand the general structure of quantum, 
and even classical and statistical, dynamics.
STA has been mainly applied to the quantum systems with 
small degrees of freedom~\cite{STA},
but, in principle, it is also possible to apply to large systems such as AQC.

The structure of this paper is as follows.
We first review STA in Sec.~\ref{sta}.
Then, we apply the method to the Grover's problem~\cite{Grover, FG}.
The Grover's problem is one of the fundamental models of AQC~\cite{RC}.
We consider two possible applications: quantum adiabatic brachistochrone 
from counterdiabatic driving (Sec.~\ref{qab}) and
inverse engineering based on the Lewis--Riesenfeld invariant (Sec.~\ref{inveng}).
We finally summarize the results and discuss future perspectives in
Sec.~\ref{summary}.

%%%%%%%%%%%%%%%%%%%%%%%%%%%%%%%%%%%%%%%%%%%%%%%%%%%%%%%%%%%%%%%%%%%%%%%%%%%%%%%%%
\section{Shortcuts to adiabaticity}
\label{sta}

%%%%%%%%%%%%%%%%%%%%%%%%%%%%%%%%%%%%%%%%%%%%%%%%%%%%%%%%%%%%%%%%%%%%%%%%%%%%%%%%%
\subsection{A crash course on shortcuts to adiabaticity}

We discuss the theoretical aspects of STA. 
There are several ways to implement STA to dynamical systems.
Although the method is best characterized theoretically by 
the Lewis--Riesenfeld invariant,
we discuss the counterdiabatic driving before that.
The counterdiabatic driving is also called the assisted adiabatic passage
or the transitionless quantum driving.
The Lewis--Riesenfeld invariant was originally proposed to solve systems 
with the time-dependent harmonic oscillator potential in 1969~\cite{LR}.
The counterdiabatic driving was proposed independently by several groups in 
the first decades of the 2000s~\cite{EZAN, DR1, DR2, Berry}.
The developments of the relation to the Lewis--Riesenfeld invariant and 
the applications to quantum control started in 2010~\cite{CRSCGM}.
Since then we can find various developments, 
some of them are described below, and a lot of experimental 
implementations~\cite{SSVL, SSCVL, Betal, Zetal, ALdCK, ZJS, NSFS} in literature.

We treat time-dependent systems with the Hamiltonian $\hat{H}_0(t)$.
This includes the AQC Hamiltonian in Eq.~(\ref{haqc}).
If the Hamiltonian is varied sufficiently slowly,
the state follows the instantaneous eigenstates of the Hamiltonian.
In systems with finite speed, we observe nonadiabatic transitions.
Then, we find that the probability to obtain the ground state 
becomes smaller than unity.
Especially, when there exists the anti-crossing of energy levels,
we have the Landau--Zener transition between 
the adjacent levels~\cite{Landau, Zener}.

The nonadiabatic transitions are suppressed 
by introducing an additional term to the Hamiltonian~\cite{EZAN, DR1, DR2, Berry}.
The Schr\"odinger equation is written as 
\be
 i\frac{\partial}{\partial t}|\psi_{\rm ad}(t)\rangle 
 =\left(\hat{H}_0(t)+\hat{H}_{\rm CD}(t)\right)|\psi_{\rm ad}(t)\rangle,
\ee
where $|\psi_{\rm ad}(t)\rangle$ represents the adiabatic state of $\hat{H}_0(t)$.
$\hat{H}_{\rm CD}(t)$ is the additional term called the counterdiabatic term.
$|\psi_{\rm ad}(t)\rangle$ is an approximate solution 
of the Schr\"odinger equation with the Hamiltonian $\hat{H}_0(t)$.
Here, the equation becomes exact by introducing the counterdiabatic term.
The adiabatic condition is not required any more.

To obtain the explicit form of the counterdiabatic term,
we need to know the detailed structure of $\hat{H}_0(t)$.
$\hat{H}_0(t)$ is formally expressed by the spectral representation as 
\be
 \hat{H}_0(t)=\sum_n \epsilon_n(t)|n(t)\rangle\langle n(t)|.
\ee
Then, the adiabatic state is given by 
\be
 |\psi_{\rm ad}(t)\rangle &=& \sum_n c_n\exp\left[
 -i\int_{0}^t \diff t'\,\epsilon_n(t')\right] \no\\
 &&\times\exp\left[-\int_{0}^t \diff t'\,\langle n(t')|\dot{n}(t')\rangle
  \right]|n(t)\rangle, \label{psiad}
\ee
where the dot denotes the time derivative.
$\{c_n\}$ is a set of constants determined by the initial condition at $t=0$.
In this adiabatic state, the probability that the state is in 
one of the eigenstates $|n(t)\rangle$, given by $|c_n|^2$, 
is independent of $t$.
By differentiating $|\psi_{\rm ad}(t)\rangle$ with respect to $t$,
we can obtain the formal expression of the counterdiabatic term:
\be
 \hat{H}_{\rm CD}(t)= i\sum_{n}\left(1-|n(t)\rangle\langle n(t)|\right)
 |\dot{n}(t)\rangle\langle n(t)|. \label{cd}
\ee
This operator has an offdiagonal form when we represent the matrix 
by the instantaneous eigenstate basis $\{|n(t)\rangle\}$.
As we mentioned above, this term prevents nonadiabatic transitions.
This term is strongly related to the adiabatic theorem as we see from 
a different representation of the counterdiabatic term: 
\be
 \hat{H}_{\rm CD}(t)= i\sum_{m\ne n}|m(t)\rangle
 \frac{\langle m(t)|\partial_t\hat{H}_0(t)|n(t)\rangle}
 {\epsilon_n(t)-\epsilon_m(t)}
 \langle n(t)|. \label{cd2} 
\ee
We note that the naive version of the adiabatic condition is written as 
\be
 \frac{|\langle m(t)|\partial_t\hat{H}_0(t)|n(t)\rangle|}
 {(\epsilon_n(t)-\epsilon_m(t))^2}\ll 1. \label{adcond}
\ee

For a given $\hat{H}_0(t)$, we solve the eigenstate equation and the solution 
is used to construct the counterdiabatic term added to the Hamiltonian.
Then, we obtain the ``adiabatic'' time evolution 
for the original Hamiltonian $\hat{H}_0(t)$.
This is not an approximation.
Once if we can obtain the counterdiabatic term,
we can  realize the adiabatic state evolution with arbitrary speed.
We note that $|\psi_{\rm ad}(t)\rangle$ is not the adiabatic state of 
the total Hamiltonian $\hat{H}(t)=\hat{H}_0(t)+\hat{H}_{\rm CD}(t)$
but of $\hat{H}_0(t)$.
In this sense, we have a nonadiabatic time evolution with respect to $\hat{H}(t)$.

We note that the counterdiabatic term had been used to describe 
the theoretical aspect of the adiabatic 
approximation~\cite{Kato, Messiah, ASY, KNS}, 
before the development of STA.
STA showed that the counterdiabatic term is useful not only 
for the formal analysis but also for practical applications.

The counterdiabatic driving can be characterized theoretically 
by the Lewis--Riesenfeld invariant~\cite{LR}.
It is a Hermitian operator $\hat{F}(t)$ satisfying the relation 
\be
 i\partial_t\hat{F}(t)=[\hat{H}(t),\hat{F}(t)] \label{lreq}
\ee
for a given Hamiltonian $\hat{H}(t)$.
This equation has the same form as the von Neumann equation.
In that case, $\hat{F}(t)$ represents the density operator.
We also mention that the Floquet operator 
$\hat{H}(t)-i\partial_t$ for periodic systems 
is interpreted as the Lewis--Riesenfeld invariant, 
if the Hilbert space where the time-derivative operator acts is defined properly. 
The Lewis--Riesenfeld invariant is not necessarily positive operator.
By using Eq.~(\ref{lreq}), we can show the following three properties:
(i). The eigenvalues of $\hat{F}(t)$ are independent of $t$:
\be
 \hat{F}(t)=\sum_n f_n |n(t)\rangle\langle n(t)|.
\ee
(ii). The solution  of the Schr\"odinger equation 
is written as Eq.~(\ref{psiad}).
$\{|n(t)\rangle\}$ represents the eigenstates of $\hat{F}(t)$
and the absolute values of the coefficients are independent of $t$.
(iii). The Hamiltonian is divided into two parts as 
$\hat{H}(t)=\hat{H}_0(t)+\hat{H}_{\rm CD}(t)$.
$\hat{H}_0(t)$ represents an operator that commutes with $\hat{F}(t)$,
and $\hat{H}_{\rm CD}(t)$ is expressed as Eq.~(\ref{cd}).

If we can find the invariant, the state can be obtained by solving 
the eigenvalue problem.
In Sec.~\ref{inveng}, we discuss 
how this method is implemented to the quantum control problem.
In this formulation, we do not introduce additional terms to the original 
Hamiltonian.
Rather, the Hamiltonian is separated into two parts.
We note that this separation is generally possible as we can understand from
the existence of the solution of Eq.~(\ref{lreq}),  
which implies that 
any quantum dynamics can be understood by the picture of 
the counterdiabatic driving.

It is well known in general quantum systems that 
the Hamiltonian plays two important roles: measure of the system energy 
and generator of the time evolution.
The present picture shows that the energy is measured by $\hat{H}_0(t)$ and
the state evolution is achieved by the generator $\hat{H}_{\rm CD}(t)$.
The state is not changed 
by the time evolution operator $\exp(-i\Delta t\hat{H}_0(t))$
since the operator is diagonal in the instantaneous basis.
It only affects the phase.
On the other hand, $\exp(-i\Delta t\hat{H}_{\rm CD}(t))$ changes 
the state to a different one.
Thus, the time evolution operator is represented by 
these two kinds of operators as 
\be
 && {\rm T}\exp\left(-i\int \diff t\,\hat{H}(t)\right) \no\\
 &=&\e^{-i\Delta t\hat{H}_{\rm CD}(t)}\e^{-i\Delta t\hat{H}_0(t)}
 \e^{-i\Delta t\hat{H}_{\rm CD}(t-\Delta t)}\e^{-i\Delta t\hat{H}_0(t-\Delta t)}\cdots,\no\\
\ee
where ${\rm T}$ denotes the time ordering and $\Delta t$ represents
an infinitesimal time interval.

The equation for the Lewis--Riesenfeld invariant 
also appears in the method of quantum brachistochrone~\cite{CHKO06, CHKO07}.
For a given constraint, an operator $\hat{F}(t)$ is defined and 
the quantum brachistochrone equation is given by Eq.~(\ref{lreq}),
which show that the optimal path is characterized by STA~\cite{kt13-2}.
Using this formulation, we can also study the stability of 
the counterdiabatic driving~\cite{kt13-2}.

%%%%%%%%%%%%%%%%%%%%%%%%%%%%%%%%%%%%%%%%%%%%%%%%%%%%%%%%%%%%%%%%%%%%%%%%%%%%%%%%%
\subsection{Examples of shortcuts to adiabaticity}

We show two examples that can obtain the counterdiabatic term explicitly.
First we consider the single-spin Hamiltonian 
\be
 \hat{H}_0(t)=h(t)\bm{n}(t)\cdot\hat{\bm{S}},
\ee
where $h(t)$ represents the magnitude of the magnetic field 
applied to the spin, $\bm{n}(t)$ is the unit vector representing 
the direction of the magnetic field.
The spin operator $\hat{\bm{S}}=(\hat{S}_1,\hat{S}_2,\hat{S}_3)$
satisfies the commutation relation 
\be
 [\hat{S}_i,\hat{S}_j]=i\epsilon_{ijk}\hat{S}_k. \label{spin}
\ee
Then, the counterdiabatic term is calculated 
as~\cite{EZAN, DR1, DR2, Berry, CLRGM} 
\be
 \hat{H}_{\rm CD}(t)=\bm{n}(t)\times\dot{\bm{n}}(t)\cdot\hat{\bm{S}}.
\ee
This example clearly indicates the basic concept of STA.
If we consider the magnetic field rotating in $xy$ plane,
the direction of the magnetic field in the counterdiabatic term 
is in $z$ direction.
Quantum fluctuation effects coming from the commutation relation 
in Eq.~(\ref{spin}) prevent the spin from staying in the $xy$ plane.
The counterdiabatic term suppresses unwanted fluctuations of the spin.
We note that the counterdiabatic term is determined by $\bm{n}(t)$ 
and is independent of $h(t)$.
This is because the change of $h(t)$ does not induce nonadiabatic transitions.

It is instructive to see that the counterdiabatic term 
introduces an operator which is not present in the original Hamiltonian.
The standard AQC uses the Ising model 
in a transverse field in $x$ direction.
The counterdiabatic driving for the single spin systems implies that
the fluctuations inevitably require additional operators in the Hamiltonian.
In other words, the counterdiabatic term for 
the stoquastic Hamiltonian~\cite{BDOT} is a nonstoquastic one.

The second example is described by the Hamiltonian
\be
 \hat{H}_0(t)=\frac{1}{2m}\hat{p}^2+\frac{1}{r^2(t)}
 U\left(\frac{\hat{x}-x_0(t)}{r(t)}\right),
\ee
where $\hat{x}$ is the position operator and $\hat{p}$ is the momentum operator.
$U$ represents an arbitrary potential function.
The time dependence comes from the dilation $r(t)$ and the translation $x_0(t)$.
This examples is known as the scale-invariant 
systems~\cite{Jarzynski, delCampo, DJC}.
Using the property that the potential function has a single argument,
we can calculate the counterdiabatic term explicitly. 
We have 
\be
 \hat{H}_{\rm CD}(t)= 
 \frac{\dot{r}(t)}{2r(t)}\left[\left(\hat{x}-x_0(t)\right)\hat{p}
 +\hat{p}\left(\hat{x}-x_0(t)\right)\right]
 +\dot{x}_0(t)\hat{p}.
\ee
This form was first obtained for the harmonic oscillator potential~\cite{MCILR}.
This is first order in $\hat{p}$ and can be represented in a form with 
the gauge potential.
The counterdiabatic term represents an electric field for a charged particle.

Correspondingly, the Lewis--Riesenfeld invariant can be found in 
these examples~\cite{LR, LL}.
We show the case of the two level system in Sec.~\ref{inveng}.

Most of experiments so far used these results.
In the second example, the form of the potential is given by 
the Harmonic oscillator.

%%%%%%%%%%%%%%%%%%%%%%%%%%%%%%%%%%%%%%%%%%%%%%%%%%%%%%%%%%%%%%%%%%%%%%%%%%%%%%%%%
\subsection{More on shortcuts to adiabaticity}

Here we discuss various achievements developed so far.
We expect that some of methods described below will be useful for AQC.

{\it State-dependent driving}. 
The counterdiabatic term in Eq.~(\ref{cd}) works for 
arbitrary choices of the initial condition of the state.
In practical calculations, we are mostly interested in controlling 
the ground state, for example.
When we treat the $n$th state, we can use a modified counterdiabatic term 
\be
 \hat{H}_{\rm CD}^{(n)}(t) = i\left(1-|n(t)\rangle\langle n(t)|\right)
 |\dot{n}(t)\rangle\langle n(t)|+(\mbox{h.c.}),
\ee
which means that the irrelevant terms can be dropped from the counterdiabatic term.
We have some arbitrariness when we implement STA.
Then, we can simplify the form of the counterdiabatic term.
In addition, since we are mostly not interested in the overall phase 
of the state, we can use unitary transformations to 
modify the Hamiltonian~\cite{ICTMR, MTCM, kt13-1, kt15}.

As a related method, a quantum state evolution is accelerated by using 
the fast-forward scaling~\cite{MN08, MN10, MN11, TMRM, kt14}. 
A state-dependent acceleration potential is introduced in this method.
The advantage of this method is that the operator form of 
the potential can be specified by ourselves, 
which is different from the counterdiabatic driving.
However, the method sometimes fails to find the potential~\cite{kt14}.
This can be understood from a simple spin example.
Suppose that we want to control the spin 
by using the magnetic field in $z$ direction.
This control does not work for the spin in the $z$ direction.
The spin cannot deviate from the $z$ axis by the $z$ magnetic field.

{\it Approximating the counterdiabatic term}.
There are many studies replacing the counterdiabatic term 
to a simple and realizable form
approximately~\cite{dCRZ, kt13-1, OM, Damski, MGCON, MMF, kt17-1, SP, Hatomura, OJV, HM, HL, PDMW}.
In many-body systems, the counterdiabatic term usually involves
many-body interaction terms, as we describe below.
It is approximated  by a noninteraction term.
Probably this is the most practical way to implement STA to AQC.
The problem is that the approximation depends on the method to use
and there are no guarantee that the approximation always works.
We need to study many examples to clarify what kind of properties
are important to improve the results.

{\it Many-body systems}.
In AQC, the Hamiltonian is interpreted as that for 
interacting quantum spin systems.
Various methods are invented in spin systems, and we can exploit 
such methods for STA.
It is well known that the one-dimensional XY spin Hamiltonian can be solved 
by mapping the spin system to a noninteracting fermion system~\cite{JW, LSM}.
The Hamiltonian is represented in a bilinear form of the fermion operators 
and the counterdiabatic term is obtained easily.
The problem in this case is that the form of the counterdiabatic term
is too complicated to realize.
It is represented by infinite series of many-body nonlocal interaction 
terms~\cite{dCRZ, kt13-1, Damski}.
In addition, the counterdiabatic term goes to infinity 
at the quantum phase transition point as we can understand from Eq.~(\ref{cd}).
What we can do is to use approximations such as truncating 
the series~\cite{dCRZ, Damski}, 
restricting to the ground state~\cite{kt13-1}, and so on.

{\it Relation to nonlinear integrable systems}. 
We showed two examples in which the explicit form of 
the counterdiabatic term is obtained.
As a matter of fact, the counterdiabatic term can be obtained analytically in
infinite series of Hamiltonians.
In the classical nonlinear integrable systems,
it is well known that the Lax formalism represents the integrability
of the system~\cite{Lax}.
In the Lax formalism, a pair of operators characterizes the system.
Two operators satisfy the Lax equation which has the same form as 
Eq.~(\ref{lreq}).
This means that by knowing the Lax pair we can obtain  
the corresponding counterdiabatic Hamiltonian~\cite{OT16}.
We have infinite series of the Lax pair in integrable systems 
such as the KdV hierarchy~\cite{Lax}. 
The corresponding quantum Hamiltonian is complicated 
with higher-order terms in the momentum operator, 
but by using some procedures such as restricting to the ground state, 
we can obtain a realizable Hamiltonian~\cite{OT16}.
We can also use the Toda hierarchy to solve 
the one-dimensional isotropic XY spin model.
The correspondence to the integrable systems may not be useful for practical 
applications but it is instructive to know solvable systems.

{\it Classical system}.
STA is not a specific method to the quantum systems.
We can also formulate STA for the classical systems
by using the adiabatic invariant~\cite{Jarzynski, PJ, OT17}.
Although the adiabatic theorem in classical mechanics looks very different 
from that in quantum mechanics, 
the applications of STA indicate that they are closely related with each other.

We can formulate the classical STA by using 
the Hamilton--Jacobi theory~\cite{OT17}.
In STA, the Hamiltonian is separated into two parts.
Correspondingly, the Hamilton--Jacobi equation is also separated into two parts.
The new generalized action defined in the Hamilton--Jacobi formalism 
can be a key quantity to find the quantum--classical correspondence.

We can also consider the Lax formalism for classical systems.
The dispersionless limit of the KdV equation is known in integrable 
systems~\cite{Lebedev, Zakharov}.
The classical limit corresponds to the dispersionless limit
and the commutator in the Lax equation~(\ref{lreq}) is replaced 
by the Poisson bracket.
Using this correspondence, 
we can find infinite series of dispersionless KdV hierarchy 
and the corresponding counterdiabatic driving in classical systems.

{\it Geometric meaning of the counterdiabatic term}.
When the Hamiltonian is written by a set of time-dependent parameters 
$\bm{\lambda}(t)=(\lambda_1(t),\lambda_2(t),\dots)$ 
as $\hat{H}_0=\hat{H}_0(\bm{\lambda(t)})$, 
the counterdiabatic term is written as 
\be
 \hat{H}_{\rm CD}(t)=\dot{\bm{\lambda}}(t)\cdot\hat{\bm{\xi}}(\bm{\lambda}(t)).
\ee
$\hat{\xi}_i(\bm{\lambda})$ represents the counterdiabatic term 
for variation of parameter $\lambda_i$.
This means that the counterdiabatic term $\hat{\xi}_i(\bm{\lambda})$
represents the generator for the parameter $\lambda_i$~\cite{Jarzynski}.
When we decompose the counterdiabatic term as above, we can show that 
a pair of the counterdiabatic terms $(\hat{\xi}_j,\hat{\xi}_k)$ 
satisfies the zero curvature condition
\be
 i\partial_{\lambda_j} \hat{\xi}_k(\bm{\lambda})
 -i\partial_{\lambda_k} \hat{\xi}_j(\bm{\lambda})
 =[\hat{\xi}_j(\bm{\lambda}),\hat{\xi}_k(\bm{\lambda})].
\ee
This clearly indicates the geometric role of 
the counterdiabatic term~\cite{SYCPS, NT}.
This equation can be useful to obtain the counterdiabatic term.
We can also consider the deformation of the integration path 
in $(t,\bm{\lambda})$ plane~\cite{NT}.

{\it Quantum speed limit and energetic cost}.
In the Mandelstam--Tamm relation~\cite{MT}, the energy 
variance $\Delta E=\sqrt{\langle \hat{H}^2\rangle-\langle \hat{H}\rangle^2}$ 
plays the role of velocity for the state evolution,
which is known as the quantum speed limit~\cite{DC}.
It is also interpreted as the energy cost and is used 
to study optimal control of the system~\cite{SS, CSHS, ZCCP, CD}.
If we implement the counterdiabatic driving the energy cost is represented 
by the counterdiabatic term: 
\be
 \Delta E(t)= \sqrt{\langle\psi_{\rm ad}(t)|
 \hat{H}^2_{\rm CD}(t)|\psi_{\rm ad}(t)\rangle}.
\ee
By using this relation, we can study an optimization of AQC.
A related study is done in the next section.

{\it Statistical dynamics}.
Although AQC treats closed systems, 
the effects of the coupling to the environment cannot be ignored 
in the realistic quantum annealing devices.
There are several works to study thermal effects by using STA.
The initial state is prepared by the canonical distributions
and we consider the time evolution in closed system.
Then, it was shown that the work fluctuation is characterized 
by the counterdiabatic term~\cite{FZCKUdC} and 
the entropy production is separated, again, into two parts~\cite{kt17-2}.
We can also apply the idea of STA to 
the Master equation~\cite{TO} and the stochastic equations.
The stochastic equations have a similar form to the Schr\"odinger equation and 
it is not difficult in principle to apply the idea of STA to such systems.

%%%%%%%%%%%%%%%%%%%%%%%%%%%%%%%%%%%%%%%%%%%%%%%%%%%%%%%%%%%%%%%%%%%%%%%%%%%%%%%%%
\section{Quantum adiabatic brachistochrone for Grover's problem}
\label{qab}

In this section, we treat the Grover's problem as a demonstration of STA.
As a possible application, we consider an optimization of the schedule
by using the method of quantum adiabatic brachistochrone.
The availability of STA is on the choice of the error function. 

%%%%%%%%%%%%%%%%%%%%%%%%%%%%%%%%%%%%%%%%%%%%%%%%%%%%%%%%%%%%%%%%%%%%%%%%%%%%%%%%%
\subsection{Grover Hamiltonian and the counterdiabatic driving}

In the Grover's search problem, 
we want to find the marked state $|0\rangle$ 
among $N$ states $|0\rangle, |1\rangle, \ldots, |N-1\rangle$.
The oracle knows the solution and we repeat queries until the solution is obtained.
Classically, the queries take $N$ steps in average.
The quantum algorithm outperforms the classical one and 
we find quadratic speedup $\sqrt{N}$~\cite{Grover, FG}.

To implement the problem by AQC, we consider the Hamiltonian~\cite{FGGS, RC} 
\be
 \hat{H}(t)= 
 A(t)\left(1-|+\rangle\langle +|\right)
 +B(t)\left(1-|0\rangle\langle 0|\right),
\ee
where 
\be
 |+\rangle = \frac{1}{\sqrt{N}}\sum_{i=0}^{N-1}|i\rangle.
\ee
Starting from the initial Hamiltonian 
$\hat{H}(0)=A(0)(1-|+\rangle\langle +|)$
with the initial ground state $|\psi(0)\rangle=|+\rangle$, 
we consider the time evolution with the Hamiltonian $\hat{H}(t)$.
$A(t)$ is monotonically decreasing from $A(0)$$(>0)$ to 0
and $B(t)$ increasing from 0 to $B(t_{\rm f})$$(>0)$.
After the time evolution, the Hamiltonian 
is given by $\hat{H}(t_{\rm f})=B(t_{\rm f})(1-|0\rangle\langle 0|)$.
If the Hamiltonian varies sufficiently slowly,
the final state is expected to be $|\psi(t_{\rm f})\rangle \sim |0\rangle$. 
This adiabatic approximation works at large $t_{\rm f}$.

Our Hamiltonian can be effectively expressed 
in two-dimensional Hilbert space.
We set the basis by using $|0\rangle$ and 
\be
 |\phi\rangle = \frac{1}{\sqrt{N-1}}\sum_{i=1}^{N-1}|i\rangle.
\ee
Then, the Hamiltonian is represented in the two-dimensional space as 
\be
 \hat{H}(t) = E_0(t)\hat{I}_2+\frac{1}{2}\Delta (t)\bm{n}(t)\cdot\hat{\bm{\sigma}},
 \label{2dH}
\ee
where $\hat{\bm{\sigma}}=(\hat{\sigma}_x,\hat{\sigma}_y,\hat{\sigma}_z)$ 
is the Pauli operator vector and 
\be
 && E_0(t) = \frac{1}{2}(A(t)+B(t)), \\
 && \Delta(t) = \sqrt{\left(A(t)-B(t)\right)^2+\frac{4}{N}A(t)B(t)}, \\
 && \bm{n}(t) = \left(-\sin\theta(t), 0, \cos\theta(t)\right), \\
 && \tan\theta(t) 
 = \frac{2\frac{\sqrt{N-1}}{N}A(t)}{\left(1-\frac{2}{N}\right)A(t)-B(t)}.
\ee
The instantaneous Hamiltonian 
has eigenvalues $E_0(t)\pm \frac{\Delta(t)}{2}$.
We note that $\Delta(t)$ represents the energy gap 
between the two eigenstates.

For this Hamiltonian, the counterdiabatic term is calculated as 
\be
 \hat{H}_{\rm CD}(t) 
 = \frac{i}{2}\dot{\theta}(t)
 \left(|0\rangle\langle\phi|-|\phi\rangle\langle 0|\right),
\ee
where
\be
 \dot{\theta}(t)=2\frac{\sqrt{N-1}}{N}\frac{A(t)\dot{B}(t)-B(t)\dot{A}(t)}
 {\Delta^2(t)}.
\ee
It is difficult to implement this Hamiltonian 
without knowing the marked state $|0\rangle$.
This is a natural result since 
the counterdiabatic driving works only when we know where to go.
Below, we utilize the result to optimize the schedule $(A(t), B(t))$, 
which can be done without implementing the counterdiabatic term.

%%%%%%%%%%%%%%%%%%%%%%%%%%%%%%%%%%%%%%%%%%%%%%%%%%%%%%%%%%%%%%%%%%%%%%%%%%%%%%%%%
\subsection{Quantum adiabatic brachistochrone}

We consider an optimization of the schedule for a fixed $t_{\rm f}$.
AQC works when adiabatic condition is satisfied.
Referring to the adiabatic condition in Eq.~(\ref{adcond}), 
an error function (``Lagrangian'') is defined as~\cite{RKHLZ, RALZ} 
\be
 L_{\rm QAB} = \frac{\Tr (\partial_t\hat{H}(t))^2}{\Delta^{4}(t)}
 =\frac{\dot{A}^2+\dot{B}^2+\frac{2}{N}\dot{A}\dot{B}}
 {\left[(A-B)^2+\frac{4}{N}AB\right]^{2}}. \label{Lqab}
\ee
The total error (``action'') is represented by 
the time integration as $S=\int \diff t\,L_{\rm QAB}$ 
and the schedule, time dependence of $(A(t),B(t))$, 
is optimized by the Euler--Lagrange equation.
Here, we propose to use the error function 
\be
 L_{\rm CD} = \frac{\Tr (\hat{H}_{\rm CD}(t))^2}
 {\Delta^{2}(t)}=\frac{\left(A\dot{B}-B\dot{A}\right)^2}
 {\left[(A-B)^2+\frac{4}{N}AB\right]^3},
\ee
instead of using Eq.~(\ref{Lqab}).
The difference is discussed below.
We note that a similar error function was discussed in Ref.~\citen{RALZ}.

By putting the parameters as $\bm{x}(t)=(A(t),B(t))$,
we can write the error functions as 
\be
 L =  \sum_{\mu,\nu}\dot{x}_\mu\dot{x}_\nu g_{\mu\nu}(\bm{x}),
\ee
which defines the metric $g_{\mu\nu}(\bm{x})$.
The introduction of the metric induces the Riemannian geometry 
and the Euler--Lagrange equation is interpreted as 
the geodesic equation~\cite{RKHLZ}.
This is applied to $L_{\rm QAB}$~\cite{RKHLZ}, but, 
in the case of $L_{\rm CD}$, the metric does not have the inverse 
and we cannot apply the geometric interpretation.
This is because two Euler--Lagrange equations are not independent with each other.
For the error function $L_{\rm CD}$, 
the equations are written as 
\be
 && B(A\ddot{B}-B\ddot{A})g_{AB} = -(A\dot{B}-B\dot{A}) \no\\
 && \times\left[
 2\dot{B}g_{AB}+\frac{1}{2}(A\dot{B}+B\dot{A})\partial_Ag_{AB}
 +B\dot{B}\partial_B g_{AB} \right], \\
 && A(A\ddot{B}-B\ddot{A})g_{AB} = -(A\dot{B}-B\dot{A}) \no\\
 && \times\left[
 2\dot{A}g_{AB}+A\dot{A}\partial_A g_{AB}
 +\frac{1}{2}(A\dot{B}+B\dot{A})\partial_Bg_{AB} \right].
\ee
These equations are combined to give 
\be
 \frac{(A\dot{B}-B\dot{A})^2}
 {\left[(A-B)^2+\frac{4}{N}AB\right]^{3}} =0.
\ee
This equation has the solution $A\dot{B}-B\dot{A} = 0$,
which describes a trivial situation~\cite{kt13-1} 
\be
 \frac{A(t)}{B(t)}={\rm const}..
\ee
In this case, the counterdiabatic term is shown to be zero.
However, this schedule is not compatible with the present 
boundary condition.
We conclude that the optimization of $L_{\rm CD}$ 
by the Euler--Lagrange equation does not work in this case.
In principle, it is still possible to minimize the error $L_{\rm CD}$
for a given boundary condition,
but the solution cannot be found from the extremization condition.
Instead, we solve the Euler--Lagrange equation by imposing some conditions
on $(A(t),B(t))$.
We show below that the method works under some constraints.

For the two-dimensional Hamiltonian in Eq.~(\ref{2dH}), 
$L_{\rm QAB}$ is written as 
\be
 L_{\rm QAB}=\frac{2\dot{E}_0^2(t)}{\Delta^4(t)}
 +\frac{\dot{\Delta}^2(t)}{2\Delta^4(t)}
 +\frac{\dot{\bm{n}}^2(t)}{2\Delta^2(t)}. \label{Lqab0}
\ee
Each term comes from the time dependence of 
$E_0$, $\Delta$, and $\bm{n}$ respectively.
We note that the last term corresponds to $L_{\rm CD}$: 
\be
 L_{\rm CD}=\frac{\dot{\bm{n}}^2(t)}{2\Delta^2(t)}.
\ee
These expressions clearly show the difference between the two error functions.
The time dependence of $E_0(t)$ does not change the state.
It only affects the overall phase and does not conflict 
with the adiabatic approximation even if $E_0(t)$ changes rapidly.
We also see that the time dependence of $\Delta(t)$ is harmless.
The change of $\Delta$ does not induce the change of the eigenstates 
when $\Delta>0$.
Thus, we consider that $L_{\rm CD}$ is more appropriate than $L_{\rm QAB}$
as an error function.

To compare the results, we consider the optimization under 
two possible constraints:
\be
 \ba{ll}
 ({\rm i}).\ \mbox{Linear constraint}
 & A(t)+B(t)=1 \\
 ({\rm ii}).\ \mbox{Quadratic constraint} & A^2(t)+B^2(t)=1
 \ea\no
\ee
For the linear constraint, 
we parametrize $A(t)=1-s(t)$ 
with the boundary conditions $s(0)=0$ and $s(t_{\rm f})=1$.
The error functions
$L_{\rm QAB}$ and $L_{\rm CD}$ are calculated as 
\be
 L \propto \left\{\ba{ll} 
 \frac{\dot{s}^2}{\left[(1-2s(t))^2+\frac{4}{N}s(t)(1-s(t))\right]^2} & 
 \mbox{QAB} \\
 \frac{\dot{s}^2}{\left[(1-2s(t))^2+\frac{4}{N}s(t)(1-s(t))\right]^3} & \mbox{CD} 
 \ea\right..
\ee
We see that the difference is in the power index of the denominator.
The Euler--Lagrange equations can be solved analytically to give~\cite{RKHLZ} 
\be
 s(t) = \left\{\ba{ll}
 \frac{1}{2}\left[1-\frac{\tan \left[(1-2\tau)\arctan\sqrt{N-1}\right]}
 {\sqrt{N-1}}\right]
 & \mbox{QAB} \\
 \frac{1}{2}\left[1-\frac{1-2\tau}{\sqrt{(1-2\tau)^2+4N\tau(1-\tau)}}\right]
 & \mbox{CD} 
 \ea\right.,
\ee
where $\tau=t/t_{\rm f}$.
For the quadratic constraint, we put 
$A(t)=\cos\varphi(t)$, $B(t)=\sin\varphi(t)$ 
with $\varphi(0)=0$ and $\varphi(t_{\rm f})=\pi/2$.
The error functions are calculated as
\be
 L \propto \left\{\ba{ll} 
 \frac{\dot{\varphi}^2(t)\left(1-\frac{1}{N}\sin 2\varphi(t)\right)}
 {\left[1-\left(1-\frac{2}{N}\right)\sin 2\varphi(t)\right]^{2}} & \mbox{QAB} \\
 \frac{\dot{\varphi}^2(t)}
 {\left[1-\left(1-\frac{2}{N}\right)\sin 2\varphi(t)\right]^3}
 & \mbox{CD} 
 \ea\right..
\ee
The schedule is obtained by solving 
\be
 \frac{t}{t_{\rm f}} = 
 \frac{\int_0^{\varphi(t)} \diff \varphi\,\sqrt{g(\varphi)}}
 {\int_0^{\pi/2} \diff \varphi\,\sqrt{g(\varphi)}},
\ee
where 
\be
 g(\varphi) = \left\{\ba{ll}
 \frac{1-\frac{1}{N}\sin 2\varphi}
 {\left[1-\left(1-\frac{2}{N}\right)\sin 2\varphi\right]^2}
 & \mbox{QAB} \\
 \frac{1}
 {\left[1-\left(1-\frac{2}{N}\right)\sin 2\varphi\right]^3} 
 & \mbox{CD} 
 \ea\right..
\ee

\begin{figure}
\includegraphics[width=1.\columnwidth]{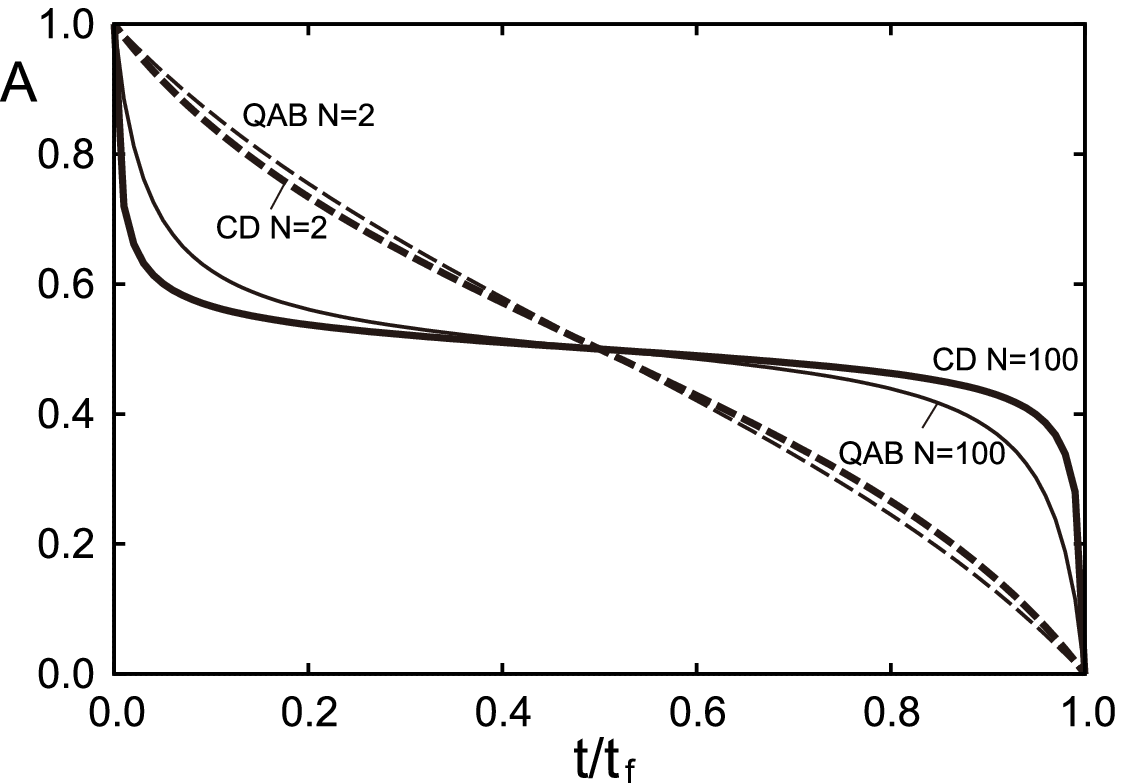}
\includegraphics[width=1.\columnwidth]{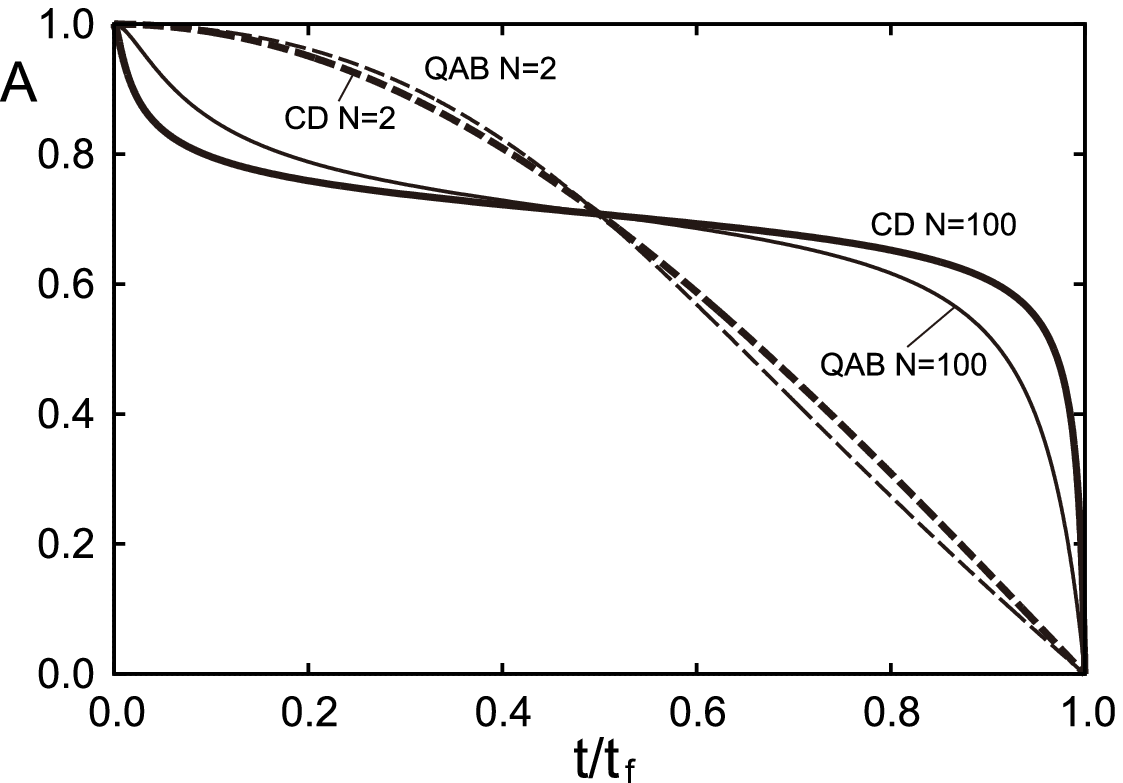}
\caption{Schedules obtained by the linear constraint $A+B=1$ (top) 
and the quadratic constraint $A^2+B^2=1$ (bottom).
``QAB'' represents the optimization by $L_{\rm QAB}$
and ``CD'' by $L_{\rm CD}$.}
\label{f1}
\end{figure}

\begin{figure}
\includegraphics[width=1.\columnwidth]{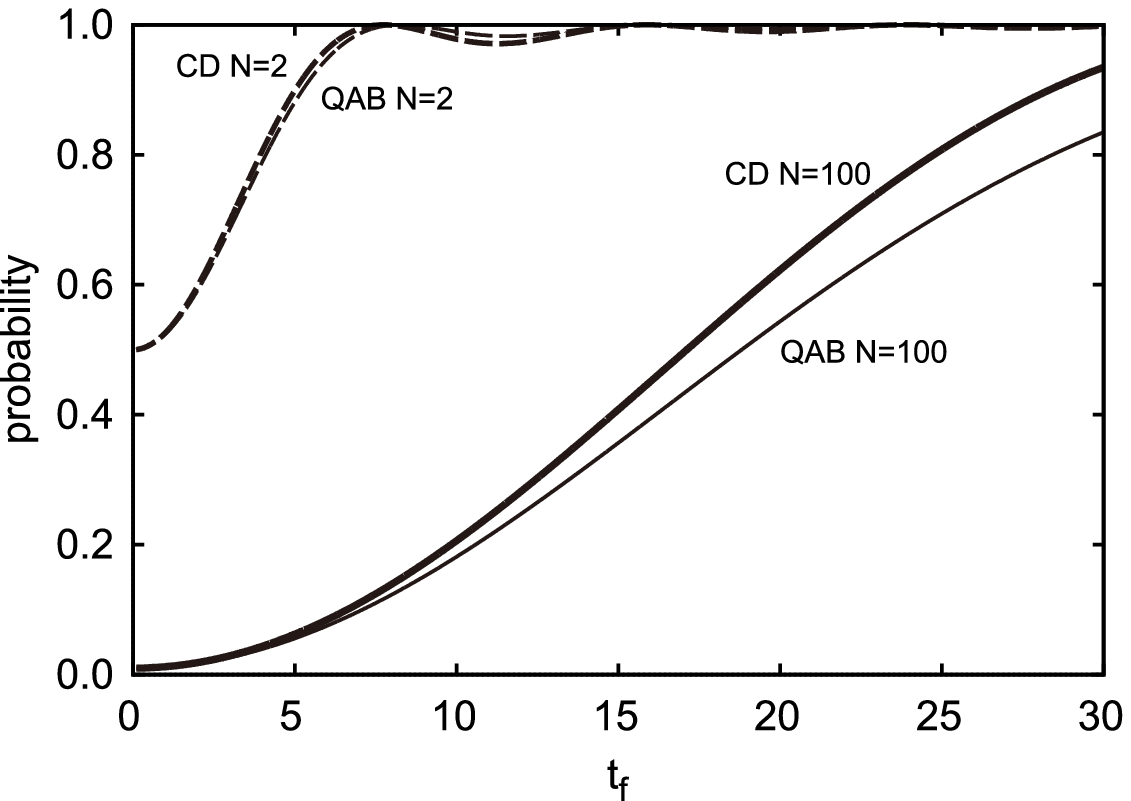}
\includegraphics[width=1.\columnwidth]{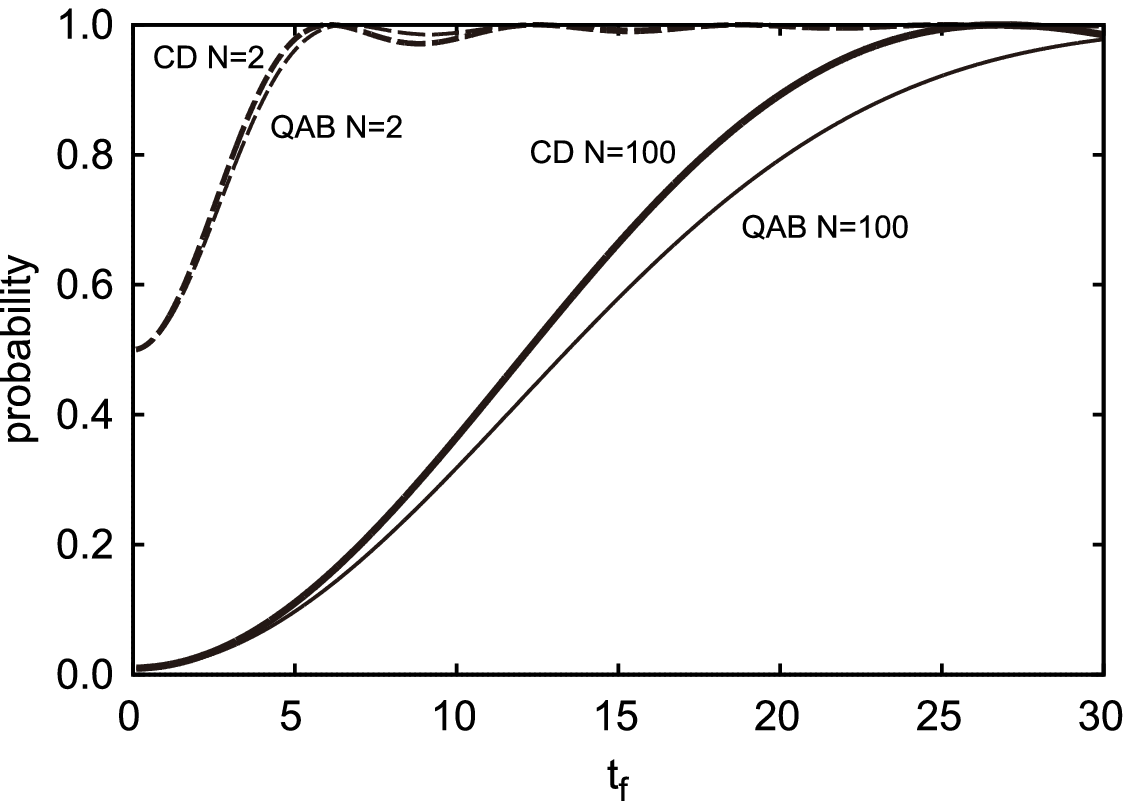}
\caption{Ground-state probability at $t=t_{\rm f}$ for 
the linear constraint $A+B=1$ (top) 
and the quadratic constraint $A^2+B^2=1$ (bottom).
``QAB'' represents the optimization by $L_{\rm QAB}$
and ``CD'' by $L_{\rm CD}$.}
\label{f2}
\end{figure}

We show the results of the schedules in Fig.~\ref{f1} and 
the corresponding results of the ground-state probability at $t=t_{\rm f}$ 
in Fig.~\ref{f2}.
We see that the $L_{\rm CD}$-optimization gives a more flat behavior of schedules 
than the $L_{\rm QAB}$-optimization 
around the intermediate time $t\sim t_{\rm f}/2$ where the energy gap becomes small.
Figure~\ref{f2} shows that $L_{\rm CD}$-optimization slightly improves the 
$L_{\rm QAB}$-optimization.
We also find that the quadratic constraint gives a better performance 
than the linear constraint. 
Although the exact reason is not clear, 
it may be related to the property that 
the quadratic constraint gives a slower change of $\Delta(t)$, 
which reduces the cost such as Eq.~(\ref{Lqab0}).

In conclusion of this section, the optimization of the schedule using STA 
may be useful but further studies are required to find the advantage.

%%%%%%%%%%%%%%%%%%%%%%%%%%%%%%%%%%%%%%%%%%%%%%%%%%%
%%%%%%%%%%%%%%%%%%%%%%%%%%%%%%%%%%%%%%%%%%%%%%%%%%%
%%%%%%%%%%%%%%%%%%%%%%%%%%%%%%%%%%%%%%%%%%%%%%%%%%%
%%%%%%%%%%%%%%%%%%%%%%%%%%%%%%%%%%%%%%%%%%%%%%%%%%%

%%%%%%%%%%%%%%%%%%%%%%%%%%%%%%%%%%%%%%%%%%%%%%%%%%%%%%%%%%%%%%%%%%%%%%%%%%%%%%%%%
\section{Lewis--Riesenfeld invariant-based inverse engineering}
\label{inveng}

Next, we discuss the control by using the Lewis--Riesenfeld invariant.
The advantage of this method is that we do not need to modify 
the original Hamiltonian.
Although we need to find the Lewis--Riesenfeld invariant 
defined in Eq.~(\ref{lreq}),
it is not necessary to solve the differential equation.
The schedule in the Hamiltonian is designed for a given solution trajectory.
The method is called the invariant-based inverse engineering~\cite{CRSCGM, STA}.

\begin{figure}
\includegraphics[width=1.\columnwidth]{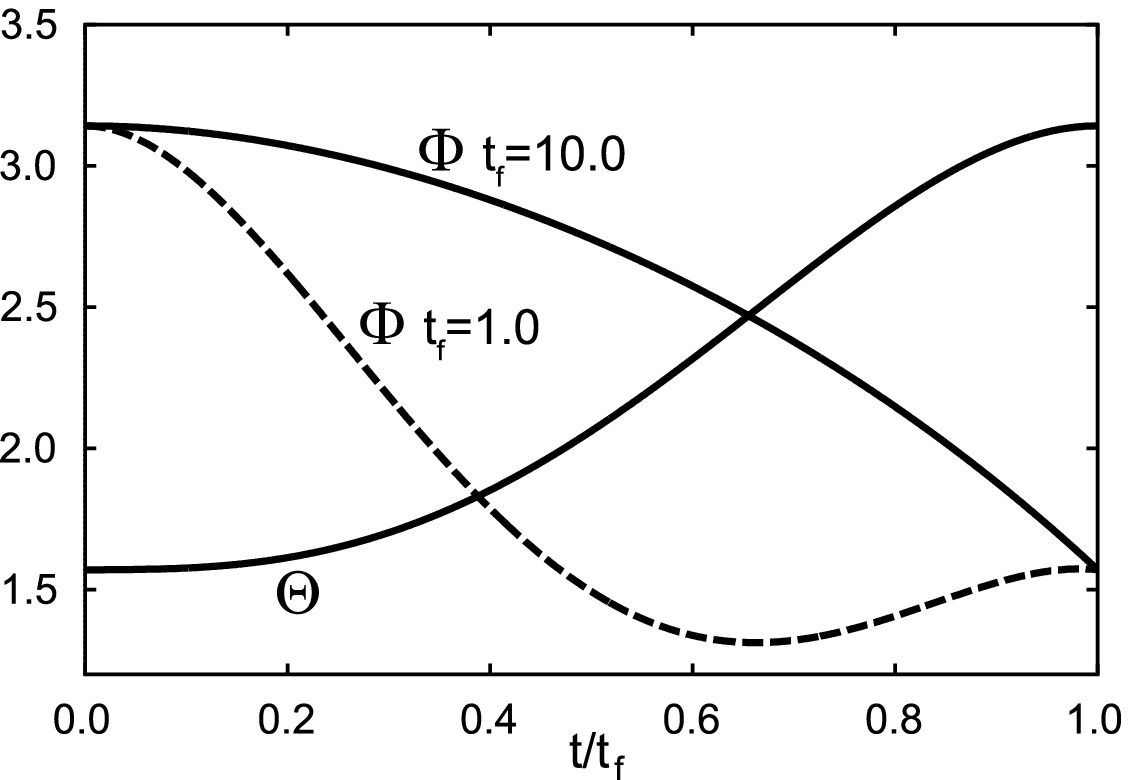}
\includegraphics[width=1.\columnwidth]{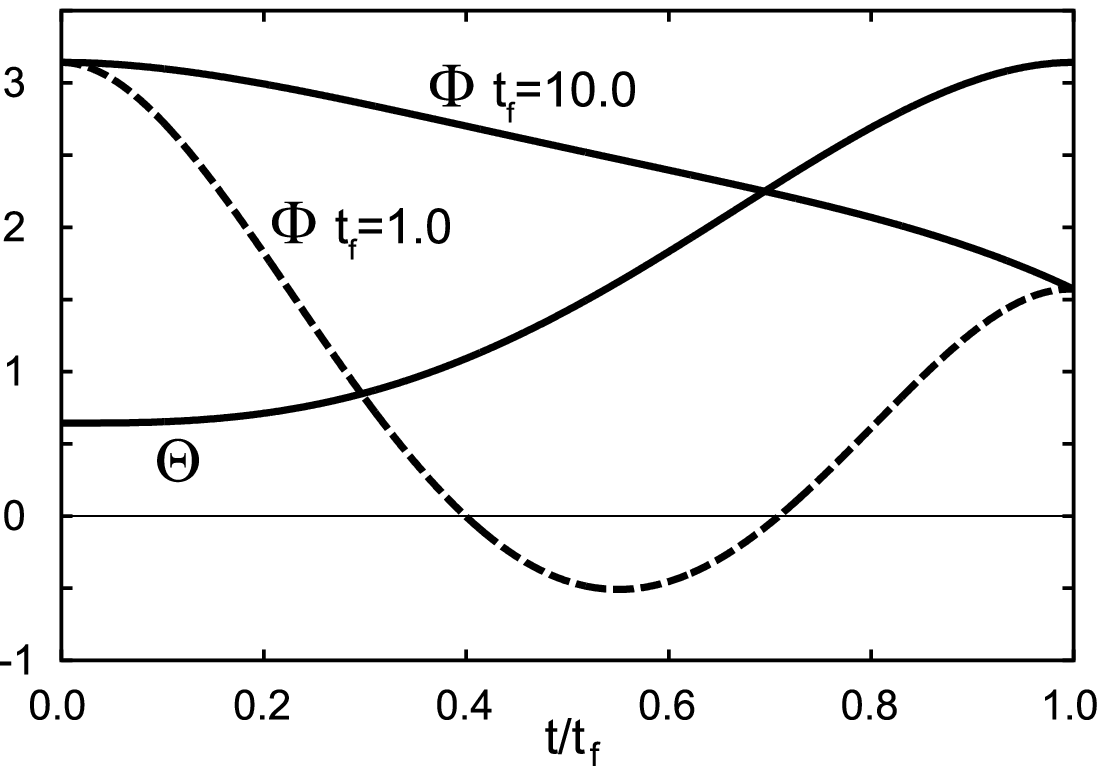}
\caption{Equations~(\ref{s1}) and (\ref{s2}) for $N=2$ (top) and $N=10$ (bottom).}
\label{f3}
\end{figure}

\begin{figure}
\includegraphics[width=1.\columnwidth]{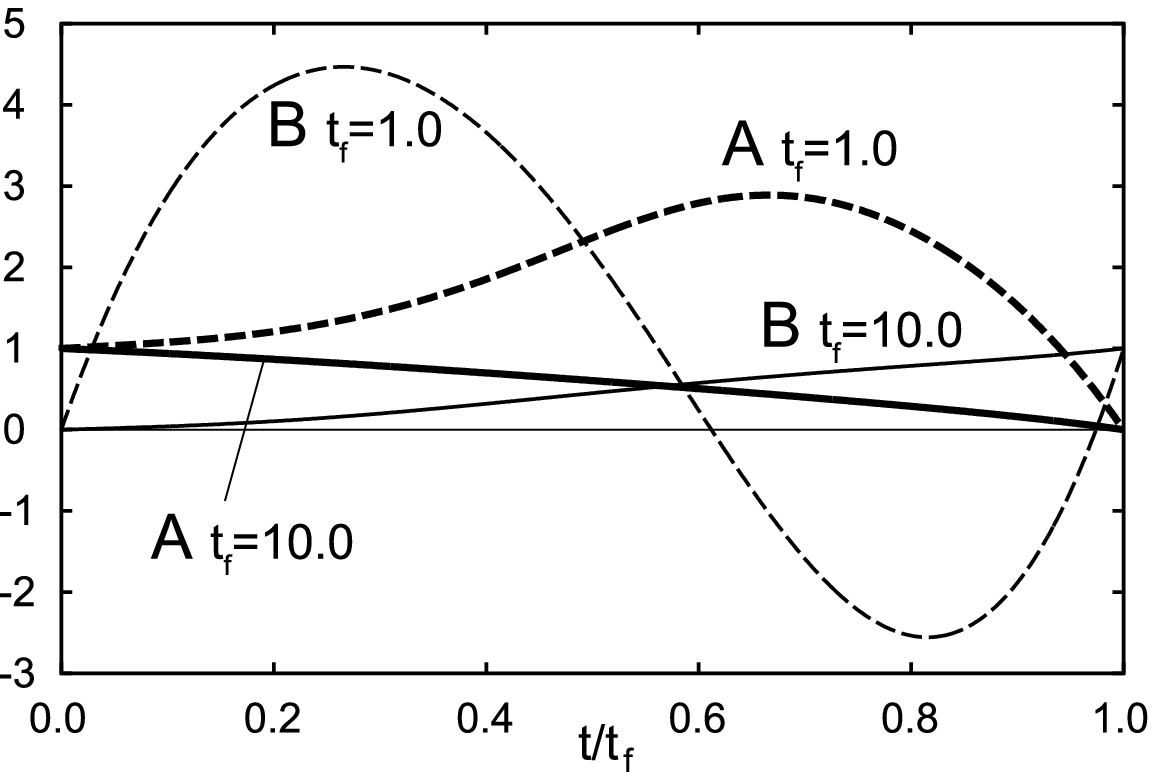}
\includegraphics[width=1.\columnwidth]{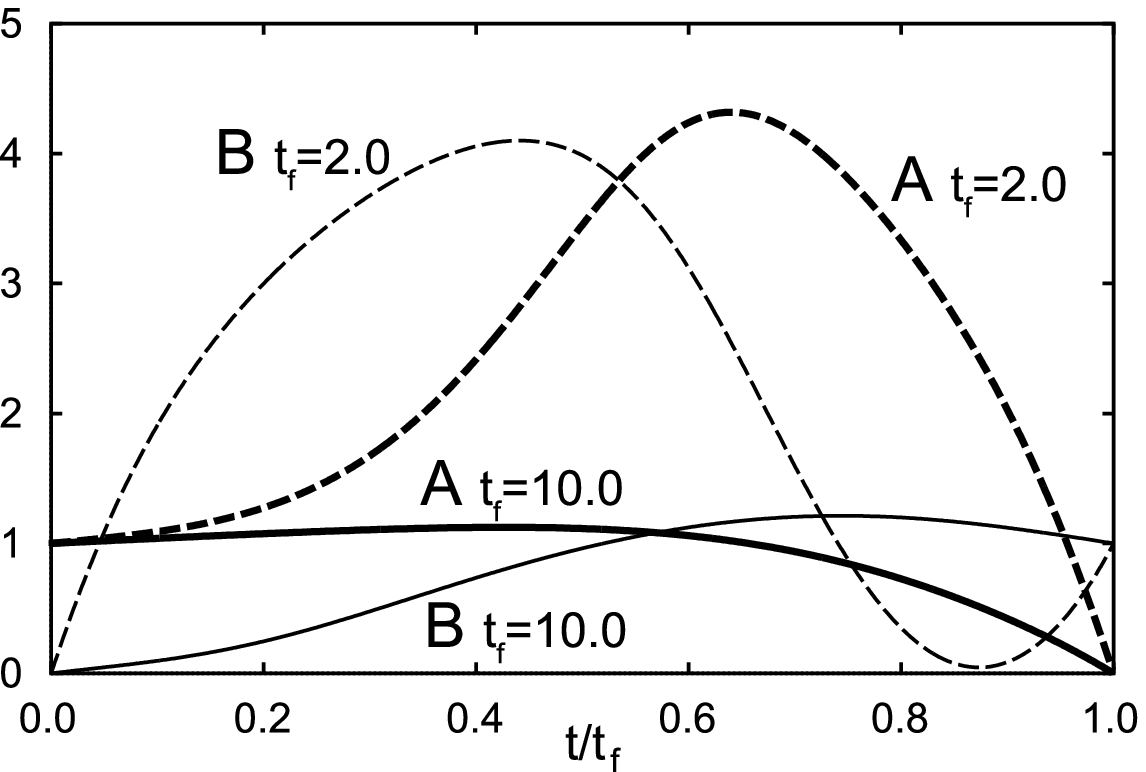}
\caption{Schedules determined by the inverse engineering in 
Eqs.~(\ref{s1}) and (\ref{s2}) for $N=2$ (top) and $N=10$ (bottom).
At $t_{\rm f}=1.0$, $A$ and $B$ are divergent at some point and are not plotted.}
\label{f4}
\end{figure}

We demonstrate the inverse engineering by using the Grover Hamiltonian 
in Eq.~(\ref{2dH}).
In the two-level system, the number of the independent Hermitian operators 
is three (except the identity operator) and 
it is not difficult to solve Eq.~(\ref{lreq}) at least in the operator level.
We put
\be
 \hat{F}(t)=\bm{e}(t)\cdot\hat{\bm{\sigma}},
\ee
where $\bm{e}(t)$ is a unit vector.
This operator has time-independent eigenvalues $\pm 1$ and we obtain 
the equation for the Lewis--Riesenfeld invariant: 
\be
 \dot{\bm{e}}(t) = \Delta(t)\bm{n}(t)\times\bm{e}(t).
\ee
In the inverse engineering, we design $\Delta(t)\bm{n}(t)$ by choosing $\bm{e}(t)$
in a proper way, which means that 
we do not need to solve the differential equation.
$\bm{e}(t)$ can be chosen arbitrary except the boundary conditions
at $t=0$ and $t=t_{\rm f}$.
At initial and final times, the state is expected to be one of 
the eigenstates of the Hamiltonian.
We require the condition
\be
 [\hat{H}(0), \hat{F}(0)]= [\hat{H}(t_{\rm f}), \hat{F}(t_{\rm f})]=0.
\ee
We parametrize the unit vector $\bm{e}(t)$
\be
 \bm{e}(t)=\left(\sin\Theta(t)\cos\Phi(t), \sin\Theta(t)\sin\Phi(t), 
 \cos\Theta(t) \right),
\ee
to write the initial condition 
\be
 \Theta(0)=\theta(0), \ \sin\Phi(0)=0,\ \dot{\Theta}(0)=0,
\ee
and the final condition
\be
 \sin\Theta(t_{\rm f})=0,\ \dot{\Theta}(t_{\rm f})=0.
\ee
For a given $(\Theta(t),\Phi(t))$, the schedule is obtained as 
\be
 && A(t)=\frac{N}{2\sqrt{N-1}}\frac{\dot{\Theta}(t)}{\sin\Phi(t)}, \label{Ainv}\\
 && B(t)=\left(1-\frac{2}{N}\right)A(t)
 +\frac{\dot{\Theta}(t)}{\tan\Theta(t)\tan\Phi(t)}
 -\dot{\Phi}(t). \no\\
\ee
In the standard procedure, we use a polynomial function 
to parametrize $(\Theta(t),\Phi(t))$~\cite{CTM, kt17-1}.
For example, a possible form is given by 
\be
 && \Theta(t)=\theta(0)\left(1-4\tau^3+3\tau^4\right)+4\pi\tau^3-3\pi\tau^4, 
 \label{s1}\\
 && \Phi(t)=\pi\left(1-2\tau^3+\frac{3}{2}\tau^4\right)
 +\frac{t_{\rm f}}{3}\left(\tau^3-\tau^4\right) \no\\
 && 
 +\frac{6N(\theta(0)-\pi)}{\sqrt{N-1}t_{\rm f}}\left(\tau^2-2\tau^3+\tau^4\right), 
 \label{s2}
\ee
where $\tau=t/t_{\rm f}$.
We plot these functions in Fig.~\ref{f3} 
and the corresponding schedule in Fig.~\ref{f4}.
Using the schedule obtained from these functions, we can realize 
the ideal time evolution.
In principle, the final state at $t=t_{\rm f}$ is exactly equal to 
the ground state of the problem Hamiltonian.
We note that the state is not in an eigenstate of the Hamiltonian 
at intermediate values of $t$.

%%%%%%%%%%%%%%%%%%%%%%%%%%%%%%%%%%%%%%%%%%%%%%%%%%%
%%%%%%%%%%%%%%%%%%%%%%%%%%%%%%%%%%%%%%%%%%%%%%%%%%%
%%%%%%%%%%%%%%%%%%%%%%%%%%%%%%%%%%%%%%%%%%%%%%%%%%%
%%%%%%%%%%%%%%%%%%%%%%%%%%%%%%%%%%%%%%%%%%%%%%%%%%%

There are several problems in this method.
For example, for a given $(\Theta(t),\Phi(t))$, 
$A(t)$ and $B(t)$ sometimes become very large or, even worse, are divergent.
In fact, for $(\Theta(t),\Phi(t))$ in Eqs.~(\ref{s1}) and (\ref{s2}),
$A(t)$ and $B(t)$ are divergent at large $N$ and small $t_{\rm f}$, 
Thus, although 
this method should work well in principle, 
there are some difficulties for the practical applications.

%%%%%%%%%%%%%%%%%%%%%%%%%%%%%%%%%%%%%%%%%%%%%%%%%%%%%%%%%%%%%%%%%%%%%%%%%%%%%%%%%
\section{Summary and perspectives}
\label{summary}

We have reviewed STA and discussed applications of STA 
to the Grover's problem.
There are two implementations of STA to dynamical systems.
We summarize the methods and discuss their advantages and problems.

{\it Counterdiabatic driving}: 
\begin{itemize}
\item This method realizes the ``adiabatic'' time evolution 
of the original Hamiltonian by introducing the additional counterdiabatic term.
The idea is simple: the additional term suppresses the nonadiabatic transitions.

\item Two typical problems are: (i). the general form of 
the counterdiabatic term is written in the spectral representation 
and it is often difficult to obtain the operator form explicitly.
(ii). Even if the counterdiabatic term is obtained theoretically,
it is too complicated to realize in experiments.

\item When the counterdiabatic term is difficult to find,
we can use some approximation methods to replace the exact form 
of the counterdiabatic term to a simple and realizable form.
We can expect partial suppression of the nonadiabatic transitions
by using such methods.

\item Originally, the counterdiabatic term was introduced to measure 
how much the time-evolution state deviates from the ideal adiabatic state.
In this paper, we used such an idea to optimize 
the schedule in the Grover Hamiltonian. 
There is no need to realize the counterdiabatic term in this method.
We can expect that the method becomes a general strategy 
to determine the schedule in AQC.

\item Possible forms of the counterdiabatic term 
can be a guiding principle to determine new driver Hamiltonians 
for future quantum annealing machines.
For the transverse Ising model, 
the explicit form of the counterdiabatic term has not been obtained.
We can only say that the counterdiabatic term 
involves the operator $\hat{\sigma}^y$ such as 
the magnetic field in $y$ direction and 
two-body interaction terms such as 
$\hat{\sigma}_i^x\hat{\sigma}_j^y$ and $\hat{\sigma}_i^z\hat{\sigma}_j^y$.

\end{itemize}

{\it Lewis--Riesenfeld invariant-based inverse engineering}: 
\begin{itemize}
\item By solving the equation 
for the Lewis--Riesenfeld invariant 
for a given Hamiltonian, we can realize an ideal time evolution 
of the state.

\item There is no need to introduce an additional term to the Hamiltonian.
We can determine the time dependence of the coefficients in the Hamiltonian.

\item Since the original form of the Hamiltonian is unchanged,  
it is very convenient for the quantum control problem.
In this case, the state follows an adiabatic passage,  
denoted by the Lewis--Riesenfeld invariant, 
which is different from the passage, 
denoted by the original Hamiltonian.

\item Generally, it is not so simple to solve the equation for 
the Lewis--Riesenfeld invariant.
However, we should stress that 
we do not need to solve the differential equation,
which is an advantage of this method.

\item The solution is not unique.
We determine the solution by requiring that it satisfies the boundary conditions.
It sometimes gives a large value as we see in Fig.~\ref{f4},
especially for small $t_{\rm f}$.

\end{itemize}

In STA, we need to know the adiabatic state of the system throughout the 
time evolution.
However, AQC is a method to find the unknown final state.
Then, it is impossible to implement STA to such systems.
STA implies that 
any system can be understood from the picture of the adiabatic 
time evolution.
The adiabatic state to obtain is different from the naive
adiabatic state.
It is important to know the difference between the two adiabatic states.
There are still some questions and problems in STA.
We expect that we can find unexpected use of STA in future studies 
to solve the optimization problem using AQC.

\begin{acknowledgment}
%\acknowledgment
This work was supported by JSPS KAKENHI Grant Number JP26400385. 
\end{acknowledgment}

%%%%%%%%%%%%%%%%%%%%%%%%%%%%%%%%%%%%%%%%%%%%%%%%%%%%%%%%%%%%%%%%%%%%%%%%%%%%%%%%%


\begin{thebibliography}{9}

\bibitem{ACF}
B.~Apolloni, C.~Carvalho, and D.~de~Falco, 
Stoch. Proc. Appl. {\bf 33}, 233 (1989).

\bibitem{Somorjai}
R.~L.~Somorjai, J. Phys. Chem. {\bf 95}, 4141 (1991).

\bibitem{AHS}
P.~Amara, D.~Hsu, and J.~E.~Straub, 
J. Phys. Chem. {\bf 97}, 6715 (1993).

\bibitem{FGSSD}
A.~B.~Finnila, M.~A.~Gomez, C.~Sebenik, C.~Stenson, and J.~D.~Doll,
Chem. Phys. Lett. {\bf 219}, 343 (1994).

\bibitem{KN}
T.~Kadowaki and H.~Nishimori, 
%{\it Quantum annealing in the transverse Ising model}, 
Phys. Rev. E {\bf 58}, 5355 (1998).

\bibitem{BBRA}
J.~Brooke, D.~Bitko, T.~F.~Rosenbaum, and G.~Aeppli, 
%{\it Quantum annealing of a disordered magnet}, 
Science {\bf 284}, 779 (1999).

\bibitem{FGGS}
E.~Farhi, J.~Goldstone, S.~Gutmann, and M.~Sipser, 
%{\it Quantum computation by adiabatic evolution}, 
arXiv:quant-ph/0001106.

\bibitem{FGGLLP}
E.~Farhi, J.~Goldstone, S.~Gutmann, J.~Lapan, A.~Lundgren, and D.~Preda, 
%{\it A quantum adiabatic evolution algorithm applied to random instances of an NP-complete problem}, 
Science {\bf 292}, 472 (2001).

\bibitem{BF}
M.~Born and V.~Fock, 
%Beweis des adiabatensatzes, 
Z. Phys. {\bf 51}, 165 (1928).

\bibitem{Kato}
T.~Kato, 
%On the adiabatic theorem of quantum mechanics, 
J. Phys. Soc. Jpn. {\bf 5}, 435 (1950).

\bibitem{JRS}
S.~Jansen, M.-B.~Ruskai, and R.~Seiler, 
J. Math. Phys. (N.Y.) {\bf 48}, 102111 (2007).

\bibitem{LRH}
D.~A.~Lidar, A.~T.~Rezakhani, and A.~Hamma, 
J. Math. Phys. (N.Y.) {\bf 50}, 102106 (2009).

\bibitem{AL18}
T.~Albash and D.~A.~Lidar,
%{\it Adiabatic quantum computation},
Rev. Mod. Phys. {\bf 90}, 015002 (2018).

\bibitem{STA}
E.~Torrontegui, S.~Ib\'a\~nez, S.~Mart\'inez-Garaot, M.~Modugno, 
A.~del~Campo, D.~Gu\'ery-Odelin, A.~Ruschhaupt, X.~Chen, and J.~G.~Muga, 
%{\it Shortcuts to adiabaticity}, 
Adv. At. Mol. Opt. Phys. {\bf 62}, 117 (2013).

\bibitem{Grover}
L.~K.~Grover, 
%{\it Quantum Mechanics Helps in Searching for a Needle in a Haystack},
Phys. Rev. Lett. {\bf 79}, 325 (1997).

\bibitem{FG}
E.~Farhi and S.~Gutmann
%{\it An analog analogue of a digital quantum computation},
Phys. Rev. A {\bf 57}, 2403 (1998).

\bibitem{RC}
J.~Roland and N.~J.~Cerf,
%{\it Quantum search by local adiabatic evolution},
Phys. Rev. A {\bf 65}, 042308 (2002).

\bibitem{LR}
H.~R.~Lewis and W.~B.~Riesenfeld, 
%An exact quantum theory of the time-dependent harmonic oscillator and of a charged particle in a time-dependent electromagnetic field, 
J. Math. Phys. (N.Y.) {\bf 10}, 1458 (1969).

\bibitem{EZAN}
A.~Emmanouilidou, X.-G.~Zhao, P.~Ao, and Q.~Niu, 
%{\it Steering an eigenstate to a destination}, 
Phys. Rev. Lett. {\bf 85}, 1626 (2000). 

\bibitem{DR1}
M.~Demirplak and S.~A.~Rice, 
%{\it Adiabatic population transfer with control fields}, 
J. Phys. Chem. A {\bf 107}, 9937 (2003).

\bibitem{DR2}
M.~Demirplak and S.~A.~Rice, 
%{\it Assisted adiabatic passage revisited}, 
J. Phys. Chem. B {\bf 109}, 6838 (2005).

\bibitem{Berry}
M.~V.~Berry, 
%{\it Transitionless quantum driving}, 
J. Phys. A {\bf 42}, 365303 (2009).

\bibitem{CRSCGM}
X.~Chen, A.~Ruschhaupt, S.~Schmidt, A.~del~Campo, D.~Gu\'ery-Odelin, 
and J.~G.~Muga, 
%{\it Fast Optimal Frictionless Atom Cooling in Harmonic Traps: Shortcut to Adiabaticity}, 
Phys. Rev. Lett. {\bf 104}, 063002 (2010).

\bibitem{SSVL}
J.-F.~Schaff, X.-L.~Song, P.~Vignolo, and G.~Labeyrie, 
%Fast optimal transition between two equilibrium states, 
Phys. Rev. A {\bf 82}, 033430 (2010).

\bibitem{SSCVL}
J.-F.~Schaff, X.-L.~Song, P.~Capuzzi, P.~Vignolo, and G.~Labeyrie, 
%Shortcut to adiabaticity for an interacting Bose-Einstein condensate, 
Europhys. Lett. {\bf 93}, 23001 (2011).

\bibitem{Betal}
M.~G.~Bason, M.~Viteau, N.~Malossi, P.~Huillery, E.~Arimondo, D.~Ciampini, R.~Fazio, V.~Giovannetti, R.~Mannella, and O.~Morsch, 
%High-fidelity quantum driving,
Nat. Phys. {\bf 8}, 147 (2012).

\bibitem{Zetal}
J.~Zhang, J.~H.~Shim, I.~Niemeyer, T.~Taniguchi, T.~Teraji, H.~Abe, S.~Onoda, T.~Yamamoto, T.~Ohshima, J.~Isoya, and D.~Suter, 
%Experimental Implementation of Assisted Quantum Adiabatic Passage in a Single Spin, 
Phys. Rev. Lett. {\bf 110}, 240501 (2013).

\bibitem{ALdCK}
S.~An, D.~Lv, A.~del~Campo, and K.~Kim, 
Nat. Comm. {\bf 7}, 12999 (2016).

\bibitem{ZJS}
X.~Zhou, S.~Jin, and J.~Schmiedmayer,
New J. Phys. {\bf 20}, 055005 (2018).

\bibitem{NSFS}
G.~Ness, C.~Shkedrov, Y.~Florshaim, and Y.~Sagi, 
New J. Phys. {\bf 20}, 095002 (2018).

\bibitem{Landau}
L.~Landau, 
%Zur Theorie der Energieubertragung II, 
Phys. Sov. Union {\bf 2}, 46 (1932).

\bibitem{Zener}
C.~Zener, 
%Non-adiabatic crossing of energy levels, 
Proc. Royal Soc. A {\bf 137}, 696 (1932).

\bibitem{Messiah}
A.~Messiah, {\it Quantum Mechanics} (Dover, New York, 2014) republication, p.752.

\bibitem{ASY}
J.~E.~Avron, R.~Seiler, and L.~G.~Yaffe, 
Comm. Math. Phys. {\bf 110}, 33 (1987).

\bibitem{KNS}
T.~Kashiwa, S.~Nima, and S.~Sakoda, 
Ann. Phys. {\bf 220}, 248 (1992).

\bibitem{CHKO06}
A.~Carlini, A.~Hosoya, T.~Koike, and Y.~Okudaira,
Phys. Rev. Lett. {\bf 96}, 060503 (2006).

\bibitem{CHKO07}
A.~Carlini, A.~Hosoya, T.~Koike, and Y.~Okudaira, 
Phys. Rev. A {\bf 75}, 042308 (2007).

\bibitem{kt13-2}
K.~Takahashi, J. Phys. A: Math. Theor. {\bf 46}, 315304 (2013).

\bibitem{CLRGM}
X.~Chen, I.~Lizuain, A.~Ruschhaupt, D.~Gu\'ery-Odelin, and J.~G.~Muga, 
Phys. Rev. Lett. {\bf 105}, 123003 (2010).

\bibitem{BDOT}
S.~Bravyi, D.~P.~DiVincenzo, R.~I.~Oliveira, and B.~M.~Terhal, 
Quant. Inf. Comp. {\bf 8}, 0361 (2008).

\bibitem{Jarzynski}
C.~Jarzynski, Phys. Rev. A {\bf 88}, 040101(R) (2013).

\bibitem{delCampo}
A.~del~Campo, 
%Shortcuts to Adiabaticity by Counterdiabatic Driving, 
Phys. Rev. Lett. {\bf 111}, 100502 (2013).

\bibitem{DJC}
S.~Deffner, C.~Jarzynski, and A.~del~Campo, Phys. Rev. X {\bf 4}, 021013 (2014).

\bibitem{MCILR}
J.~G.~Muga, X.~Chen, S.~Ib\'a\~nez, I.~Lizuain, and A.~Ruschhaupt, 
J. Phys. B: At. Mol. Opt. Phys. {\bf 43}, 085509 (2010).

\bibitem{LL}
H.~R.~Lewis and P.~G.~L.~Leach,
%A direct approach to finding exact invariants for one‐dimensional time‐dependent classical Hamiltonians
J. Math. Phys. {\bf 23}, 2371 (1982).

\bibitem{ICTMR}
S.~Ib\'a\~nez, X.~Chen, E.~Torrontegui, J.~G.~Muga, and A.~Ruschhaupt, 
Phys. Rev. Lett. {\bf 109}, 100403 (2012).

\bibitem{MTCM}
S.~Mart\'inez-Garaot, E.~Torrontegui, X.~Chen, and J.~G.~Muga, 
Phys. Rev. A {\bf 89}, 053408 (2014).

\bibitem{kt13-1}
K.~Takahashi, Phys. Rev. E {\bf 87}, 062117 (2013).

\bibitem{kt15}
K.~Takahashi, Phys. Rev. A {\bf 91}, 042115 (2015).

\bibitem{MN08}
S.~Masuda and K.~Nakamura, Phys. Rev. A {\bf 78}, 062108 (2008).

\bibitem{MN10}
S.~Masuda and K.~Nakamura, Proc. R. Soc. A {\bf 466}, 1135 (2010).

\bibitem{MN11}
S.~Masuda and K.~Nakamura, Phys. Rev. A {\bf 84}, 043434 (2011).

\bibitem{TMRM}
E.~Torrontegui, S.~Mart\'inez-Garaot, A.~Ruschhaupt, and J.~G.~Muga, 
Phys. Rev. A {\bf 86}, 013601 (2012).

\bibitem{kt14}
K.~Takahashi, Phys. Rev. A {\bf 89}, 042113 (2014).

\bibitem{dCRZ}
A.~del~Campo, M.~M.~Rams, and W.~H.~Zurek, 
Phys. Rev. Lett. {\bf 109}, 115703 (2012).

\bibitem{OM}
T.~Opatrn\'y and K.~M\o lmer, New J. Phys. {\bf 16},  015025 (2014).

\bibitem{Damski}
B.~Damski, J. Stat. Mech. P12019 (2014).

\bibitem{MGCON}
S.~Masuda, U.~G\"ung\"ord\"u, X.~Chen, T.~Ohmi, and M.~Nakahara, 
Phys. Rev. A {\bf 93}, 013626 (2016).

\bibitem{MMF}
V.~Mukherjee, S.~Montangero, and R.~Fazio, 
%Local shortcut to adiabaticity for quantum many-body systems, 
Phys. Rev. A {\bf 93}, 062108 (2016).

\bibitem{kt17-1}
K.~Takahashi, Phys. Rev. A {\bf 95}, 012309 (2017).

\bibitem{SP}
D.~Sels and A.~Polkovnikov, PNAS {\bf 114}, E3909 (2017).

\bibitem{Hatomura}
T.~Hatomura, J. Phys. Soc. Jpn. {\bf 86}, 094002 (2017).

\bibitem{OJV}
A.~B.~\"Ozg\"uler, R.~Joynt, and M.~G.~Vavilov, 
Phys. Rev. A {\bf 98}, 062311 (2018).

\bibitem{HM}
T.~Hatomura and T.~Mori, Phys. Rev. E {\bf 98}, 032136 (2018).

\bibitem{HL}
A.~Hartmann and W.~Lechner, arXiv:1807.02053.

\bibitem{PDMW}
F.~Petiziol, B.~Dive, F. ~Mintert, and S.~Wimberger, 
Phys. Rev. A {\bf 98}, 043436 (2018).

\bibitem{JW}
P.~Jordan and E.~Wigner, Zeits. f. Physik A {\bf 47}, 631 (1928).

\bibitem{LSM}
E.~Lieb, T.~Schultz, and D.~Mattis, Ann. Phys. (NY) {\bf 16}, 407 (1961).

\bibitem{Lax}
P.~D.~Lax, 
%Integrals of nonlinear equations of evolution and solitary waves, 
Commun. Pure Appl. Math. {\bf 21}, 467 (1968).

\bibitem{OT16}
M.~Okuyama and K.~Takahashi, Phys. Rev. Lett. {\bf 117}, 070401 (2016).

\bibitem{PJ}
A.~Patra and C.~Jarzynski, J. Phys. Chem. B  {\bf 121}, 3403 (2017).

\bibitem{OT17}
M.~Okuyama and K.~Takahashi, J. Phys. Soc. Jpn. {\bf 86}, 043002 (2017).

\bibitem{Lebedev}
D.~R.~Lebedev, Phys. Lett. A {\bf 74}, 154 (1979).

\bibitem{Zakharov}
V.~E.~Zakharov, Funct. Anal. Appl. {\bf 14}, 89 (1980).

\bibitem{SYCPS}
N.~A.~Sinitsyn, E.~A.~Yuzbashyan, V.~Y.~Chernyak, A.~Patra, and C.~Sun, 
Phys. Rev. Lett. {\bf 120}, 190402 (2018).

\bibitem{NT}
K.~Nishimura and K.~Takahashi, SciPost Phys. {\bf 5}, 029 (2018).

\bibitem{MT}
L.~Mandelstam and I.~Tamm, J. Phys. (Moscow) {\bf 9}, 249 (1945).

\bibitem{DC}
S.~Deffner and S.~Campbell,
J. Phys. A: Math. Theor. {\bf 50}, 453001 (2017).

\bibitem{SS}
A.~C.~Santos and M.~S.~Sarandy, 
Sci. Rep. {\bf 5}, 15775 (2015).

\bibitem{CSHS}
I.~B.~Coulamy, A.~C.~Santos, I.~Hen, and M.~S.~Sarandy, 
Front. ICT {\bf 3}, 19 (2016).

\bibitem{ZCCP}
Y.~Zheng, S.~Campbell, G.~De~Chiara, and D.~Poletti, 
Phys. Rev. A {\bf 94}, 042132 (2016).

\bibitem{CD}
S.~Campbell and S.~Deffner, 
Phys. Rev. Lett. {\bf 118}, 100601 (2017).

\bibitem{FZCKUdC}
K.~Funo, J.-N.~Zhang, C.~Chatou, K.~Kim, M.~Ueda, and A.~del~Campo,
% universal work fluctuations during shortcuts to adiabaticity by counterdiabatic driving, 
Phys. Rev. Lett. {\bf 118}, 100602 (2017).

\bibitem{kt17-2}
K.~Takahashi, New J. Phys. {\bf 19}, 115007 (2017).

\bibitem{TO}
K.~Takahashi and M.~Ohzeki, 
Phys. Rev. E {\bf 93}, 012129 (2016).

\bibitem{RKHLZ}
A.~T.~Rezakhani, W.~J.~Kuo, A.~Hamma, D.~A.~Lidar, and P.~Zanardi, 
%{\it Quantum Adiabatic Brachistochrone}, 
Phys. Rev. Lett. {\bf 103}, 080502 (2009).

\bibitem{RALZ}
A.~T.~Rezakhani, D.~F.~Abasto, D.~A.~Lidar, and P.~Zanardi, 
Phys. Rev. A {\bf 82}, 012321 (2010).

\bibitem{CTM}
X.~Chen, E.~Torrontegui, and J.~G.~Muga,
Phys. Rev. A {\bf 83}, 062116 (2011).

\end{thebibliography}
\end{document}